\documentclass[11pt]{article}
\usepackage[final]{acl}
\usepackage{times}
\usepackage{latexsym}
\usepackage[T1]{fontenc}
\usepackage[utf8]{inputenc}
\usepackage{microtype}
\usepackage{inconsolata}
\usepackage{algorithm}
\usepackage{algorithmic}
\usepackage{amsmath,amssymb,amsfonts}
\usepackage{booktabs}
\usepackage[table]{xcolor}
\usepackage{arydshln}
\definecolor{goodcell}{RGB}{209, 229, 240}
\definecolor{bestcell}{RGB}{44,123,182}
\definecolor{badcell}{RGB}{253, 208, 162}   \definecolor{worstcell}{RGB}{215,25,28}

\colorlet{tableheadcolor}{gray!25} 
\colorlet{tablerowcolor}{gray!10} 
\colorlet{tablerowcolor2}{gray!45} 
\colorlet{tablerowcolor3}{gray!12} 

\newcommand{\rowcollight}{\rowcolor{tablerowcolor3}} %

\usepackage{epigraph}
\usepackage{caption}
\usepackage{subcaption,graphicx}
\usepackage[table]{xcolor}
\usepackage{pgf}

\newcommand{\gradcell}[1]{%
  \begingroup
  \pgfmathsetmacro{\val}{#1}%
  \def\cellshade{}%
  \ifdim\val pt>0pt
    \pgfmathtruncatemacro{\shade}{min(80,round(80*\val/0.75))}%
    \xdef\cellshade{\noexpand\cellcolor{poscolor!\shade!white}}%
  \else
    \ifdim\val pt<0pt
    \pgfmathtruncatemacro{\shade}{min(80,round(80*abs(\val)/0.75))}%
      \xdef\cellshade{\noexpand\cellcolor{negcolor!\shade!white}}%
    \fi
  \fi
  \endgroup
  \cellshade #1%
}

\definecolor{negcolor}{HTML}{FF69B4}
\definecolor{poscolor}{HTML}{006400}

\definecolor{DarkBlue}{HTML}{00008B}
\definecolor{mscolor}{HTML}{01665e}
\definecolor{nmscolor}{HTML}{bf812d}
\definecolor{lgreen}{HTML}{ccece6}
\definecolor{dolive}{HTML}{308014}
\definecolor{purple}{HTML}{ae017e}
\definecolor{brickred}{HTML}{f03b20}

\newif{\ifhidecomments}
  \hidecommentsfalse 
\ifhidecomments
    \newcommand{\agam}[1]{}
    \newcommand{\olivia}[1]{}
    \newcommand{\eshwar}[1]{}
    \newcommand{\koustuv}[1]{}
\else
    \newcommand{\agam}[1]{\textbf{\small\sffamily{\textcolor{DarkBlue}{[#1 -- Agam]}}}}
    \newcommand{\olivia}[1]{\textbf{\small\sffamily{\textcolor{dolive}{[#1 -- Olivia]}}}}
    \newcommand{\eshwar}[1]{\textbf{\small\sffamily{\textcolor{brickred}{[#1 -- Eshwar]}}}}
    \newcommand{\koustuv}[1]{\textbf{\small\sffamily{\textcolor{purple}{[#1 -- Koustuv]}}}}
  \fi

\newcommand{\para}[1]{\vspace{0.5em}\noindent\textbf{\textit{#1}~}}

\title{\textsc{Vastu}: Language Models Struggle to Recognize Online Community Values}
\author{
 \textbf{Agam Goyal},~
 \textbf{Xianyang Zhan},~
 \textbf{Charlotte Lambert},\\
 \textbf{Koustuv Saha},~
 \textbf{Eshwar Chandrasekharan}
\\
 Siebel School of Computing and Data Science\\  University of Illinois Urbana-Champaign, Urbana, IL, USA
\\
\texttt{\{agamg2, zhan39, cjl8, ksaha2, eshwar\}@illinois.edu}
}

\begin{document}

\maketitle

\def\thefootnote{\arabic{footnote}}

\begin{abstract}
Online communities develop distinct norms for content they collectively value, yet it remains unclear whether current language models can recognize locally valued contributions in context. We formalize this as \textbf{community-conditioned preference prediction} and introduce \textsc{Vastu} (\underline{V}alue-\underline{A}ware \underline{S}ocial \underline{Tu}ning), a benchmark of 75,000 Reddit comments from 15 communities spanning Gaming, Science, Q\&A, Advice, and Politics. We evaluate four model families---prompted LLMs, LoRA-adapted SLMs, supervised encoders, and feature-based classifiers---across global, local, and context-conditioned settings. Our central finding is that parametric adaptation consistently outperforms prompting: supervised encoders reach 0.74 AUROC and fine-tuned SLMs 0.64--0.71, while the best prompted result is only 0.62. This gap is not merely quantitative---vanilla prompting yields over 80\% false-negative rates, systematically discarding content communities actually value. Conversational context narrows but does not close this divide. Together, these results suggest that local preference recognition requires community-specific training signal, not just better prompting. Our work supports future research on community-aware reward modeling, feed curation, and positive moderation.\footnote{Code for experiments and the VASTU benchmark can be found in\href{https://github.com/scuba-illinois/VASTU}{https://github.com/scuba-illinois/VASTU}.}
\end{abstract}

\section{Introduction}

Research on online communities has focused heavily on detecting harmful content such as toxicity, hate speech, and misinformation~\cite{jurgens_just_2019,vidgen-etal-2021-learning,nandi-etal-2025-psychology}. But content moderation and curation are not just punitive tasks. Communities also need systems that can recognize and surface \emph{constructive} contributions---comments that provide evidence, clarify disagreements, offer support, or exemplify valued behavior~\cite{grimmelmann_virtues_2015,wang_highlighting_2021,10.1145/3686929,jundi2025not}. However, before such systems can generate interventions, recommend exemplars to newcomers, or guide moderators, they need a detection model: one that can \textit{identify which contributions a particular community values and considers desirable.}

The key challenge is that community-valued content is not the same as generic quality, helpfulness, or non-toxicity: a comment can be constructive in the abstract yet redundant in context, or valued in one community while penalized in another~\cite{weld_what_2022,10.1145/3544548.3580644,goyal2026uncovering}. Prior work on community preferences takes two forms: \emph{characterizing} community norms by decomposing them into interpretable preference dimensions or cultural value frameworks~\cite{park2024valuescope,li2025prefpalette,zhang2026cultivating}, and \emph{aligning} model outputs with community preferences through benchmarking or preference optimization~\cite{pmlr-v235-sorensen24a,kumar-etal-2025-compo,lin2026communitybench, kim2026llumi}. We study a different capability: can models predict, for a naturally occurring human-written content in its conversational context, whether a community would approve of it? This task requires understanding real online discourse dynamics, where approval depends on community-specific norms, and has direct implications for value-aware feed reranking and positive moderation~\cite{jahanbakhsh2026value}.

The consequences of model performance on this task has direct implications for NLP research. For example, a community-conditioned preference detector can function as a reward model. Given a set of candidate responses, it can rank by predicted local approval~\cite{goyal2026language}, construct weak preference pairs for community-conditioned fine-tuning of language models~\cite{kolluri-etal-2025-finetuning}, or retrieve valued exemplars for in-context learning~\cite{chen-etal-2025-spica}. This detection problem is thus a prerequisite for building generative systems that adapt to \textit{pluralistic} community norms without collapsing them into a single generic notion of quality~\cite{feng-etal-2024-modular,zhang-etal-2025-exploring-chain}.

We address these gaps by formalizing this as \textbf{community-conditioned preference prediction}, where given a community $c$, a target comment $x$, and optional conversational context $z$ (parent context), predict whether $x$ is likely to be locally valued by $c$. We treat upvotes as behavioral signals of local community approval and aggregate preferences. To evaluate models on this task, we introduce \textsc{Vastu} (\underline{V}alue-\underline{A}ware \underline{S}ocial \underline{Tu}ning), a benchmark of 75,000 Reddit comments from 15 communities spanning Gaming, Science, Q\&A, Advice, and Politics, with community-specific approval labels and contextual variants.

Using \textsc{Vastu}, we conduct a comprehensive analysis to study three research questions:

\noindent\textbf{RQ1: How well do current NLP models perform on community-conditioned preference prediction?} We compare feature-based baselines, supervised encoders, prompted LLMs, and LoRA-adapted SLMs under shared evaluation settings.

\noindent\textbf{RQ2: Which forms of community and conversational context improve detection?} We ablate input context (comment-only, community-aware, and parent-aware variants) to test whether models benefit from local grounding beyond comment text.

\noindent\textbf{RQ3: Where do current frontier models fail?} We perform a quantitative error profiling and qualitatively develop an error taxonomy to dive deeper into shortcomings of evaluated models.

\para{Summary of Findings:} Community-conditioned preference prediction remains far from solved. Our key finding is that adaptation outperforms prompting across all model sizes: locally fine-tuned encoders reach 0.71--0.74 AUROC, fine-tuned SLMs 0.64--0.69, and the best prompted result with full conversational context is only 0.62. Beyond aggregate performance, error profiles reveal a deeper asymmetry: vanilla LLMs discard over 80\% of community-valued content, applying generic quality heuristics that miss community-specific signals such as inside references, casual humor, and direct personal experience. Prompt engineering reduces this bias substantially but cannot eliminate it. Conversational context helps across all paradigms but does not substitute for community-specific adaptation---local models consistently outperform global ones, and what predicts value in one community type is often penalized in another.
\section{Related Work}

\para{Content Values in Online Communities:} Understanding what online communities value has been approached through multiple methodological lenses. Survey-based studies have developed taxonomies of community values from user perspectives~\cite{weld_what_2022} and moderator perspectives on what behaviors they wish to encourage~\cite{10.1145/3686929}, revealing that qualities like prosociality, engagement, and content quality are commonly desired. Complementing these qualitative approaches, \citet{goyal2026uncovering} used an LLM-based framework to identify values present in highly-upvoted Reddit comments at scale, uncovering values at macro, meso, and micro scales. Research on Reddit's voting mechanisms has shown that upvotes serve as signals of community-approved discourse styles, with \citet{10.1145/3544548.3580644} demonstrating this for political rhetoric in political subreddits. Other computational approaches to measuring valued behavior have largely focused on specific attributes such as partisan animosity~\cite{doi:10.1126/science.adu5584} and prosociality~\cite{bao2021conversations}, detected using transformer-based approaches, or identifying causal drivers of community-valued content using feature-based models~\cite{goyal2026language} for proactive content moderation. Recent work decomposes these community preferences into interpretable latent dimensions~\cite{li2025prefpalette} or maps them to cultural value frameworks~\cite{zhang2026cultivating}, revealing that community preference variation is both real and structurally complex. \textit{Our work extends this line of work by comparing computational approaches for detecting valued-aligned content, evaluating whether models can learn the implicit value signals expressed through upvotes.}

\para{Detecting and Surfacing Valued Content:} Social media feed algorithms primarily optimize for engagement signals~\cite{chan2026examining}, which can amplify divisive content and misalign with user preferences~\cite{cunningham2024we,milli2025engagement}. Recent work has explored value-aligned alternatives using platform-agnostic browser extensions that intercept and rerank feeds in real-time~\cite{10.1145/3800557}. \citet{jia2024embedding} introduced societal objective functions that translate social science constructs into ranking objectives. Building on this, \citet{jahanbakhsh2026value} implemented feeds ranked by Schwartz's Basic Human Values, showing that user-designed feeds diverge substantially from engagement-driven defaults. Systems like Bonsai~\cite{10.1145/3772318.3791855} enable natural language-based feed customization, while Alexandria~\cite{kolluri2026alexandria} offers a pluralistic library of values for real-time reranking. CommunityBench~\cite{lin2026communitybench} benchmarks community-level alignment of LLMs across diverse groups and tasks, showing that current models systematically struggle with community-specific preferences; our work complements this by focusing on the recognition of valued human-generated content in context rather than AI output alignment. A common thread across this work is the need for models that can reliably detect value-relevant content at scale. \textit{Our work addresses this fundamental challenge by comprehensively comparing modeling approaches for detecting community-valued content.}
\section{The \textsc{Vastu} Benchmark}\label{sec:data}

We now introduce the evaluation setup.

\para{Communities:}
We pick 15 topically diverse Reddit communities, grouped into five domains for balanced coverage and structured analysis:
\textbf{Gaming} (\textit{r/Games}, \textit{r/2007scape}, \textit{r/DestinyTheGame});
\textbf{Science} (\textit{r/science}, \textit{r/askscience}, \textit{r/Futurology});
\textbf{Q\&A} (\textit{r/AskWomen}, \textit{r/AskHistorians}, \textit{r/IAmA});
\textbf{Advice} (\textit{r/personalfinance}, \textit{r/legaladvice}, \textit{r/DIY});
and \textbf{Politics} (\textit{r/CanadaPolitics}, \textit{r/PoliticalDiscussion}, \textit{r/socialism}).

\para{Data curation:}
We extract comments posted in these communities between May 10, 2016 and Feb 4, 2017---a period for which data exists from prior content moderation research~\cite{chandrasekharan_internets_2018,chandrasekharan_crossmod_2019,lambert_conversational_2022}. We then remove deleted/removed comments, content from deleted/banned authors, moderator-distinguished posts, likely bots (identified by a deterministic username heuristic), and very short comments ($\leq$ 5 tokens). We operationalize community approval using subreddit-specific \textit{score} (upvotes minus downvotes) quantiles: \texttt{HIGH} for comments above the 95th percentile and \texttt{LOW} for comments below the 70th percentile following prior work~\cite{goyal2026uncovering}. This intentionally prioritizes separability over full coverage, to ensure the \texttt{HIGH} comments are capturing the upper end of content the community values. Prior work has validated using Reddit score to encode community-specific preferences while being a universal signal of approval~\cite{10.1145/3544548.3580644,park2024valuescope,lambert2025does,kumar-etal-2025-compo}. For each community, we then sample an equal number of \texttt{high} and \texttt{low} comments, capped at 2{,}500 per class (i.e., at most 5{,}000 comments per subreddit), producing a balanced dataset per community.

\para{Context Variants:} Each instance always includes the target comment text. Instances can additionally be conditioned on the subreddit name and description (community-aware), the parent post title and body (post-aware), or the direct parent comment (parent-aware). We evaluate models across these conditions to measure the marginal contribution of each context type.

\para{Features:} We compute five types of feature signals: \textbf{(i) Embeddings:} Dense representations using SentenceBERT (\texttt{all-mpnet-base-v2})~\cite{reimers-gurevych-2019-sentence}. \textbf{(ii) Prosociality:} Following \citet{bao2021conversations}, we measure six attributes: politeness, agreement, supportiveness, donation, laughter, and gratitude. \textbf{(iii) Surface-level features:} We compute sentiment via the VADER compound score~\cite{hutto2014vader}, Flesch reading ease~\cite{kincaid1975derivation}, and interrogative style (fraction of sentences ending in ``?''). \textbf{(iv) Toxicity:} We use Detoxify~\cite{Detoxify}, an open-source toxicity detection model, to estimate comment toxicity. \textbf{(v) LIWC:} We apply the LIWC 2015 dictionary~\cite{pennebaker2015development} to quantify psychosocial linguistic processes and social values, complementing features (i)--(iv) with broader coverage of implicit community norms~\cite{weld2024making,park2024valuescope}.

\para{Model families and evaluation protocol:}
We evaluate four families of models: (i) traditional feature-based models; (ii) a supervised encoder fine-tuned on the task; (iii) prompted frontier LLMs evaluated across context conditions; and (iv) LoRA-adapted SLMs fine-tuned with and without community conditioning. We use community-stratified train/test splits with an 80/20 ratio and report AUROC, F1, and Brier calibration~\cite{brier1950verification}. For robustness, we evaluate one representative model per paradigm under a stricter ($q{=}0.99$) and a softer ($q{=}0.90$) \texttt{HIGH} threshold to assess sensitivity to the approval cutoff (Appendix~\S\ref{app:threshold-sensitivity}).
\section{RQ1: Performance on Community- Conditioned Preference Prediction}

We compare four model families under shared evaluation settings: feature-based classifiers, supervised encoders, prompted LLMs, and LoRA-adapted SLMs.

\subsection{Evaluated Models and Setup}

\para{Simple Baselines:} We include two off-the-shelf baselines: \textbf{(1) 1-Toxicity} inverts the Detoxify toxicity score, testing whether non-toxic content proxies for valued content; and \textbf{(2) DialogRPT}, trained by \citet{gao-etal-2020-dialogue} on Reddit data to predict comment scores, which acts as the most natural zero-shot transfer baseline for this task.

\para{Feature-Based Models:} We evaluate two feature-based classifiers using the full feature set described in Section~\ref{sec:data}: $L_2$-regularized Logistic Regression, with regularization parameter $C$ tuned via cross-validation over $\{0.01, 0.1, 1, 10, 100\}$, and XGBoost with learning rate $0.05$, $\gamma=1$, maximum depth 4, regularization $\lambda=3$ and $\alpha=1$, and early stopping after 50 rounds. Both models are evaluated with 5-fold cross-validation. A feature set ablation confirming the incremental contribution of each feature group is provided in Appendix \S\ref{app:feature-ablation}.

\para{Supervised Encoders:} We fine-tune two encoder models on the classification task: \textit{bert-base-uncased} (BERT)~\cite{devlin-etal-2019-bert} and ModernBERT~\cite{warner2025smarter}. Both share the same architecture: transformer-encoded comment representations are combined with linguistic features via learnable fusion weights, then passed through a two-layer classification head (see Appendix \S\ref{app:transformer-architecture}). Encoders are fine-tuned end-to-end for three epochs with a learning rate of $2 \times 10^{-5}$ and batch size 16.

\para{Language Model Approaches:} We evaluate language models in two paradigms---prompting frozen models and fine-tuning open-weight models.

\noindent\textbf{(1) Prompted LLMs:} We evaluate GPT-4o-mini and GPT-4o~\cite{hurst2024gpt} (standard) and GPT-5-mini~\cite{openaiIntroducingGPT5} (reasoning model; uses \textit{``medium''} reasoning effort.) under four prompting strategies: \textbf{(i) Zero-Shot:} The model receives the comment and subreddit description and predicts whether the comment falls in the top 5\% of community upvotes; \textbf{(ii) Chain-of-Thought (CoT):} The model reasons step-by-step before predicting~\cite{wei2022chain}: identifying community values, assessing comment qualities, then synthesizing a prediction; \textbf{(iii) Few-Shot ($k$=3):} Three labeled examples from the same community (training set) precede evaluation. Examples are class-balanced and drawn from different threads to prevent leakage; and \textbf{(iv) Value-Augmented (VAP):} The model receives the Schwartz Basic Human Values framework~\cite{Schwartz1992UniversalsIT,Schwartz2012RefiningTT} as an analytical lens~\cite{2026whosevalues}, assessing whether the comment's expressed values align with the community's priorities.

All models use temperature=0 to reduce variability. Full prompt templates are in Appendix \S\ref{app:llm-prompts}.

\noindent\textbf{(2) Fine-Tuned Small Language Models:} We fine-tune Gemma-3-4B-IT~\cite{team2025gemma} and Qwen3-4B-Instruct~\cite{yang2025qwen3} using 4-bit quantization and LoRA~\cite{Hu2021LoRALA} for one epoch on community-specific training splits. Models are trained to output only the classification label (\texttt{HIGH}/\texttt{LOW}), enabling direct comparison with prompted LLMs at lower inference cost.

\para{Training Scope:} For feature-based classifiers and supervised encoders, we evaluate two training scopes. \textbf{Global} models are trained on data pooled from all 15 communities and learn shared patterns of community approval. \textbf{Local} models are trained separately per community to capture community-specific norms; we report the mean performance ($\pm$std) across all 15 local models.

\subsection{Results: Feature-Based/Encoders Models}

\begin{table}[t]
\small
\sffamily
\centering
\resizebox{\columnwidth}{!}{
\begin{tabular}{llccc}
\textbf{Model} & \textbf{Scope} & AUROC\,$\uparrow$ & F1\,$\uparrow$ & Brier\,$\downarrow$ \\
\midrule
\rowcolor{blue!10}\multicolumn{5}{l}{\textbf{Feature-based Models}} \\
LogReg & Global & $0.65$ & $0.61$ & $0.23$ \\
LogReg & Local & $0.69 \pm 0.04$ & $0.64 \pm 0.03$ & $0.22 \pm 0.01$ \\
\hdashline
XGBoost & Global & $0.66$ & $0.61$ & $0.23$ \\
XGBoost & Local & $0.67 \pm 0.04$ & $0.63 \pm 0.02$ & $0.23 \pm 0.01$ \\
\rowcolor{blue!10}\multicolumn{5}{l}{\textbf{Supervised Encoders}} \\
BERT & Global & $0.70$ & $0.67$ & $0.22$ \\
BERT & Local & $0.71 \pm 0.05$ & $0.68 \pm 0.03$ & $0.23 \pm 0.02$ \\
\hdashline
ModernBERT & Global & $0.73$ & $0.70$ & $0.21$ \\
ModernBERT & Local & \cellcolor{goodcell}$0.74 \pm 0.04$ & \cellcolor{goodcell}$0.72 \pm 0.03$ & $0.21 \pm 0.02$ \\
\rowcolor{blue!10}\multicolumn{5}{l}{\textbf{Baselines}} \\
1-Toxicity & -- & \cellcolor{badcell}$0.45\pm0.05$ & \cellcolor{badcell}$0.41\pm0.04$ & \cellcolor{badcell}$0.81\pm0.06$ \\
DialogRPT & -- & $0.59\pm0.04$ & $0.55 \pm 0.06$ & \cellcolor{goodcell}$0.17\pm0.02$ \\
\bottomrule
\end{tabular}}
\caption{\textbf{Detection performance across model families.} Global models train on pooled data; Local reports the mean ($\pm$std) across community-specific models. Per-topic breakdowns are in \S\ref{app:topic-breakdowns}.\vspace{-12pt}}
\label{tab:rq1-comparison}
\end{table}

\para{Simple baselines are insufficient.} The 1-Toxicity baseline performs below chance (0.45 AUROC), demonstrating that desirability and non-toxicity are distinct constructs: content can be non-toxic yet not valued, or seem toxic yet be highly rewarded by a community. DialogRPT performs better (0.59 AUROC) and achieves strong calibration (0.17 Brier) owing to its score-prediction training objective, but still trails the simplest feature-based model by 0.06 points, indicating that Reddit-wide engagement patterns do not capture community-specific approval.

\para{Local adaptation outperforms global training:} Locally trained models outperform their global counterparts across all models (Table~\ref{tab:rq1-comparison}). The gap is largest for Logistic Regression (+0.04 AUROC) and smaller for fully fine-tuned encoders (+0.01--0.02), suggesting that model capacity can partially compensate for the absence of community-specific supervision, but not eliminate it. ModernBERT consistently outperforms BERT under both training scopes (+0.03 AUROC), achieving the strongest overall result at 0.74 AUROC. Performance is consistently highest for Politics communities and lowest for Science across all model families, highlighting that different communities have norms that models may learn with different levels of difficulty. See \autoref{tab:rq1-topics} in Appendix \S\ref{app:topic-breakdowns} for a detailed breakdown.

\para{Calibration is consistent across approaches:} We find that Brier scores are stable across evaluated models (0.22--0.24), indicating that while detection performance varies substantially, all models produce reasonably calibrated probability estimates---an important property for downstream moderation workflows where confidence thresholds guide decisions~\cite{goyal-etal-2025-momoe}.

See Appendix \S\ref{app:shap-analysis} for SHAP experiments on feature-based models for interpretability analysis.

\subsection{Results: Language Model Approaches}

Since LLMs return discrete predictions rather than probabilities, we omit Brier calibration metrics for this subsection.

\begin{table}[t]
\small
\sffamily
\centering
\resizebox{\columnwidth}{!}{
\begin{tabular}{llcc}
\textbf{Model} & \textbf{Strategy} & AUROC\,$\uparrow$ & F1\,$\uparrow$ \\
\midrule
\rowcolor{blue!10}\multicolumn{4}{l}{\textbf{Prompted LLMs}} \\
GPT-4o-mini & Zero-Shot & $0.56$ & $0.38$ \\
 & Chain-of-Thought & $0.55$ & $0.39$ \\
 & Few-Shot ($k$=3) & $0.56$ & $0.40$ \\
 & Value-Augmented & $0.55$ & $0.45$ \\
\hdashline
GPT-4o & Zero-Shot & $0.58$ & $0.41$ \\
 & Chain-of-Thought & $0.56$ & $0.40$ \\
 & Few-Shot ($k$=3) & $0.60$ & $0.42$ \\
 & Value-Augmented & $0.60$ & \cellcolor{goodcell}$0.48$ \\
\rowcolor{blue!10}\multicolumn{4}{l}{\textbf{Prompted LRM (Reasoning)}} \\
GPT-5-mini & Zero-Shot & \cellcolor{badcell}$0.53$ & \cellcolor{badcell}$0.20$ \\
 & Chain-of-Thought & $0.55$ & $0.37$ \\
 & Few-Shot ($k$=3) & \cellcolor{badcell}$0.53$ & $0.22$ \\
 & Value-Augmented & $0.57$ & $0.38$ \\
\rowcolor{blue!10}\multicolumn{4}{l}{\textbf{Fine-tuned SLMs}} \\
Gemma-3-4B-IT & LoRA + Zero-Shot & \cellcolor{goodcell}$0.65$ & $0.42$ \\
Qwen3-4B-Instruct & LoRA + Zero-Shot & $0.64$ & $0.43$ \\
\bottomrule
\end{tabular}}
\caption{\textbf{Language model performance across paradigms.} Fine-tuned SLMs substantially outperform all prompted configurations. Reasoning models perform worst overall; Chain-of-Thought provides no consistent benefit, suggesting this task requires learned community norms rather than test-time reasoning.\vspace{-12pt}}
\label{tab:llm-prompting}
\end{table}

\para{Prompted LLMs underperform supervised approaches:} Across all prompting strategies, LLMs fall well short of fine-tuned encoders and SLMs (Table~\ref{tab:llm-prompting}). The best prompted result---GPT-4o with Few-Shot or Value-Augmented prompting (0.60 AUROC)---trails even simple Logistic Regression (0.65 Global) and is 0.14 points below the best fine-tuned encoder (ModernBERT Local: 0.74). This gap is consistent across all community types (see \autoref{tab:llm-topics} in Appendix \S\ref{app:topic-breakdowns}), mirroring the supervised model ordering: Politics remains most predictable and Gaming most challenging. These results indicate that recognizing locally valued content requires learning community-specific patterns that cannot be elicited through prompting alone.

\para{Fine-tuned SLMs perform substantially better:} LoRA fine-tuning significantly narrows the gap: Gemma-3-4B-IT (0.65 AUROC) and Qwen3-4B-Instruct (0.64 AUROC) both outperform all prompted LLMs including GPT-4o, despite being approximately 100$\times$ smaller. This aligns with recent findings that fine-tuned SLMs can match or exceed LLMs for content moderation tasks~\cite{zhan-etal-2025-slm}. Fine-tuned SLMs approach feature-based model performance but still trail fully fine-tuned encoders by 0.09 points, suggesting that the combination of contextual representations and linguistic features captures signals that pure language modeling misses.

\para{Reasoning capabilities do not help:} Neither explicit reasoning prompts nor reasoning-optimized models improve performance. Chain-of-Thought fails to outperform Zero-Shot for any model, and in some cases degrades it (GPT-4o: 0.58 to 0.56 AUROC). GPT-5-mini achieves the worst overall results (0.53 AUROC, 0.20 F1 Zero-Shot), underperforming both standard LLMs and fine-tuned SLMs. This pattern suggests that predicting community-valued content is not a reasoning task amenable to chain-of-thought decomposition: it requires implicit knowledge of community norms that must be learned from examples rather than inferred at inference time.

\para{Value-Augmented prompting improves calibration but not discrimination:} Value-Augmented prompting consistently raises F1 across prompted models (e.g., GPT-4o: 0.41 to 0.48) by improving precision-recall trade-offs, likely because the Schwartz values framework helps calibrate decision thresholds. However, AUROC is unchanged, indicating no improvement in the underlying ability to rank valued over non-valued content. Structured analytical prompts improve the threshold decision but cannot substitute for supervised training on community-specific data.

\section{RQ2: Impact of Community and Conversational Context on Prediction}

We now test whether community metadata or parent-comment context helps prompted models recognize locally valued contributions. We evaluate three input conditions across all prompted models: \textit{comment-only} (no context), \textit{+community} (subreddit name and description prepended), and \textit{+community+parent} (community metadata plus the direct parent comment).

\para{Parent-comment context provides the most consistent AUROC gain:} The \texttt{+community+parent} condition yields the largest improvement for every model. For prompted LLMs, AUROC gains over comment-only range from +0.02 (GPT-5-mini: 0.53 to 0.55) to +0.04 (GPT-4o: 0.58 to 0.62). Fine-tuned SLMs show even larger gains: Gemma-3-4B improves from 0.65 to 0.71 and Qwen3-4B from 0.64 to 0.70, both +0.06. The consistency of this pattern suggests that conversational grounding from the parent comment provides crucial context that neither model parametric scale nor community descriptions alone can replicate.

\begin{table}[t]
\small
\sffamily
\centering
\resizebox{\columnwidth}{!}{
\begin{tabular}{l cc cc cc}
\textbf{Model}
  & \multicolumn{2}{c}{\textbf{comment-only}}
  & \multicolumn{2}{c}{\textbf{+community}}
  & \multicolumn{2}{c}{\textbf{+community+parent}} \\
\cmidrule(lr){2-3}\cmidrule(lr){4-5}\cmidrule(lr){6-7}
  & AUROC\,$\uparrow$ & F1\,$\uparrow$
  & AUROC\,$\uparrow$ & F1\,$\uparrow$
  & AUROC\,$\uparrow$ & F1\,$\uparrow$ \\
\midrule
\rowcolor{blue!10}\multicolumn{7}{l}{\textbf{Prompted LLMs}} \\
GPT-4o-mini & $0.56$ & $0.38$ & $0.57$ & $0.37$ & $0.59$ & $0.41$ \\
GPT-4o      & $0.58$ & $0.41$ & $0.61$ & $0.40$ & $0.62$ & $0.44$ \\
GPT-5-mini  & \cellcolor{badcell}$0.53$ & \cellcolor{badcell}$0.20$ & $0.54$ & $0.24$ & $0.55$ & $0.21$ \\
\rowcolor{blue!10}\multicolumn{7}{l}{\textbf{Fine-tuned SLMs}} \\
Gemma-3-4B  & $0.65$ & $0.42$ & $0.66$ & $0.43$ & \cellcolor{goodcell}$0.71$ & \cellcolor{goodcell}$0.49$ \\
Qwen3-4B    & $0.64$ & $0.43$ & $0.65$ & $0.41$ & $0.70$ & $0.48$ \\
\bottomrule
\end{tabular}}
\caption{\textbf{Context ablation across model families.} Community and parent comment context improves prediction accuracy. SLMs continue to show the best overall performance, with AUROC reaching upto $0.71$.\vspace{-12pt}}
\label{tab:rq2-context-ablation}
\end{table}

\para{Community description alone is inconsistent and sometimes counterproductive:} Adding community metadata without conversational grounding produces mixed results. GPT-4o-mini shows a slight F1 drop on \texttt{+community} (0.38 to 0.37) despite a marginal AUROC gain, while GPT-5-mini's F1 spikes to 0.24 on \texttt{+community} before dropping back to 0.21 on \texttt{+community+parent}. Qwen3-4B shows a similar F1 dip on \texttt{+community} (0.43 to 0.41). These inconsistencies suggest that community descriptions shift decision thresholds rather than providing reliable alignment signal, producing variable trade-offs between precision and recall across model types.

\para{Fine-tuned SLMs show a more structured context response:} For SLMs, the gain from \texttt{+community} to \texttt{+community+parent} is substantially larger than from comment-only to \texttt{+community} (approximately $6\times$ the AUROC improvement). This asymmetry is absent in prompted LLMs, where context gains are distributed more evenly. The pattern suggests that fine-tuning already encodes community-level signals parametrically, leaving parent-comment context as the primary remaining informational lever at inference time.

\para{While context helps, the adaptation gap persists for prompted models:} Even the strongest prompted condition---GPT-4o with \texttt{+community+parent} (0.62 AUROC)---remains below fine-tuned SLMs at comment-only (0.64--0.65) and well below locally trained encoders (BERT Local: 0.71, ModernBERT Local: 0.74). Fine-tuned SLMs with parent context (0.70--0.71), however, begin approaching the lower range of local encoder performance, suggesting that parametric adaptation combined with conversational grounding is the most promising direction.

\section{RQ3: LLM Error Analysis}\label{sec:error-analysis}

Recent work on content moderation has shown promising performance of language models on the punitive side of this task, i.e., removing norm-violating content~\cite{kumar2024watch,kolla2024llm,goyal-etal-2025-momoe}. However, our task captures the positive content curation side of moderation, where we see that language models perform much worse. In order to understand further how models fail, we conduct a targeted error analysis of false positives (FPs) and false negatives (FNs). This is also crucial as systematic errors can encode miscalibrated assumptions about community norms and lead to harmful ranking decisions even when overall performance might appear strong. We therefore examine misclassified instances across communities to identify recurring failure patterns. 

This analysis yields both a quantitative error profile, and a qualitative error taxonomy of distinct failure modes, which we use to diagnose when different approaches are likely to break down, and to surface concrete directions for future work, including development of hybrid systems.

\begin{figure}[t]
    \centering
    \includegraphics[width=\linewidth]{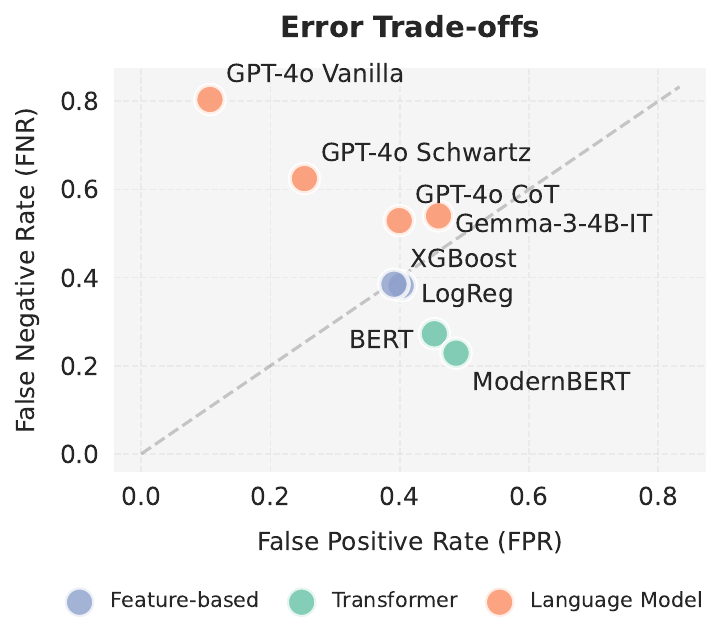}
    \caption{\textbf{Error trade-offs across method families.} While fine-tuned encoders and feature-based models tend to minimize False Negatives (prioritizing recall), vanilla LLMs show extreme conservatism in predicting valued content, resulting in high False Negative rates. Prompt engineering strategies and fine-tuning successfully shift LMs toward a more balanced error profile.\vspace{-16pt}}
    \label{fig:fp-fn-comparison}
\end{figure}

\para{Quantitative Error Profiles:} Figure~\ref{fig:fp-fn-comparison} quantifies these systematic biases by visualizing the trade-off between False Positive Rate (FPR) and False Negative Rate (FNR) for the best performing models in each category. We observe a clear divergence in how different model families approach the curation task. Supervised methods, including both feature-based and fine-tuned Transformers, cluster in the central and lower-right regions, favoring lower False Negative rates. This indicates a ``high-recall'' behavior essential for curation---they successfully surface the majority of community-valued content, at the cost of allowing some lower-quality items into the feed. In contrast, off-the-shelf LLMs, especially with vanilla prompting exhibit extreme conservatism, characterized by a really high FNR. In a curation context like ours, this behavior is highly undesired, as the model acts as a strict gatekeeper, discarding over 80\% of the content the community actually values. Notably, prompting strategies that induce reasoning or apply the Schwartz values framework drive a significant shift along the error axes, reducing False Negatives by nearly 30 percentage points. However, even with these interventions, prompted LLMs remain significantly more conservative than fine-tuned models, suggesting that without parametric adaptation, general-purpose models struggle to recognize the idiosyncratic norms of specific subreddits.

\para{Qualitative Error Taxonomy:} To better understand the mechanism behind these quantitative gaps, we manually inspected 150 misclassified examples in the \texttt{comment-only} condition for GPT-4o with CoT using thematic analysis~\cite{braun2012thematic} (10 per community), applying an inductive coding workflow. Two authors independently open-coded an initial subset of 50 explanations to identify recurring justifications (e.g., ``fits community norms,'' ``lacks context,'' ``helpful/actionable,'' ``hostile tone''). They then met to reconcile disagreements, refine and finalize a compact codebook with inclusion/exclusion criteria, and agree on higher-level themes. The first author then applied this finalized codebook to code the remaining instances. We subsequently derived a taxonomy of six distinct failure modes. 

The most prevalent error types explain the high conservatism (FNR) observed in Figure~\ref{fig:fp-fn-comparison}. Specifically, \textbf{T2: ``More effort is better''} and \textbf{T6: Profanity penalty} reveal a mismatch between the LLM's generic ``helpfulness'' prior and actual community dynamics. The model frequently rejects concise, witty, or casually profane comments that are highly valued by communities for their relatability, mistakenly flagging them as ``low effort'' or ``unprofessional.'' Similarly, \textbf{T1: Wrong community norm} highlights the difficulty of zero-shot transferability of these models. For example, in gaming communities, the model often rejected inside jokes or memes as ``irrelevant,'' failing to recognize that shared cultural references are a primary driver of value in that space. Conversely, False Positives often stemmed from \textbf{T4: Evidence bar mismatch}, where the model over-valued comments that \textit{sounded} authoritative but were factually incorrect or missed the nuance required by expert communities like \textit{r/AskHistorians}. This taxonomy underscores that while LLMs possess a strong general notion of ``quality,'' they lack the specific grounding required for community-specific curation. The other two failure modes are \textbf{T3: Missing thread context} and \textbf{T5: Humor / sarcasm}. See~\autoref{tab:rq3-error-taxonomy} in Appendix \S\ref{app:llm-error-coding} for expanded results and comparison to SHAP features for feature-based models.
\section{Discussion and Implications}

\vspace{-8pt}
\para{Beyond ``One-Size-Fits-All'' Models:} Our findings challenge the prevailing assumption that ``quality'' is a universal construct that can be captured by a single, general-purpose model. The consistently high performance of community-specific local models over global baselines demonstrates that \textit{value} is fundamentally contextual: what constitutes a high-quality contribution in \textit{r/AskHistorians} which has rigorous discussions with a neutral tone is qualitatively different from valued content in \textit{r/2007scape} which thrives on humor and community-specific jargon. For platforms, this suggests that the current paradigm of training a single reward model for platform-wide ranking may be inherently limited. Instead, we envision a shift toward decentralized and modular community-specific architectures. By deploying lightweight, community-specific adapters (e.g., LoRA modules), platforms can respect the pluralistic nature of human values~\cite{kolluri2026alexandria,pmlr-v235-sorensen24a}. Future work could formalize this using explicit routing mechanism directing content to the specific expert models that understands that community's norms, similar to systems like MoMoE~\cite{goyal-etal-2025-momoe}. Such a system would allow for scalable personalization without flattening cultural nuances central to online communities.

\para{Understanding the Limits of LLMs for Detecting Community-Valued Content:} While LLMs are increasingly being looked at as the future of content curation~\cite{2026whosevalues,jahanbakhsh2026value,goyal2026uncovering}, our error analysis reveals that current off-the-shelf LLMs act as ``moderators'' rather than ``curators.'' Vanilla LLMs exhibited prohibitively high False Negative rates ($>0.8$), systematically discarding niche or implicit content that communities actively value. This behavior likely stems from safety-tuning (RLHF) that conflates ``desirable'' with ``sanitized,'' failing to recognize that a low-effort joke or an edgy debate might be valued in specific communities.

We showed that strategies like Value-Augmented Prompting (VAP) can partially mitigate this by providing a structured lens for evaluation, improving calibration. However, we found that prompting alone cannot fully override the model's inherent conservatism and fine-tuning on community-specific data might be required. This points toward practical implications for future curation systems such as \textsc{Bonsai}~\cite{10.1145/3772318.3791855} and \textsc{Pilot}~\cite{choi2025designing}: the actual discovery mechanism---the ``engine'' of feed curation---requires high-recall models that have been parametrically adapted to community preferences, which also aligns with the idea of ``candidate generation'' in RecSys literature~\cite{covington2016deep,huang2025comprehensive}. LLMs may instead be best suited for downstream roles such as explaining why content was ranked, providing interpretable rationales, or serving as a final sanity check rather than driving initial retrieval or ranking.

\para{Towards Efficient Models:} One of the key practical findings of our work is that massive scale of models is not a prerequisite for effective value alignment. We show that fine-tuned SLMs with just 4B parameters consistently outperformed LLMs, offering a favorable trade-off between accuracy and efficiency. For developers building real-time systems---whether feed rankers, recommendation engines, or moderation tools---this is a crucial insight as this makes it computationally feasible to run much more accurate, community-aligned rankers on edge, avoiding the API cost and rate limits of calling larger, closed-source models. Furthermore, this efficiency also unlocks new possibilities for user agency. If such ``value'' detectors are small and cheap to run, platforms can move beyond opaque algorithms for feed curation and offer users a catalog of ``curation lenses'', which could allow a user to swap in a \textit{`science-focused'} ranker on their news feeds or a \textit{`support-focused'} ranker for advice or well-being communities. This would democratize the alignment process, turning curation from a platform-imposed constraint into a user-controlled utility. Future research should explore the user experience of such controls, using \textsc{Vastu} for evaluating models powering them.
\section{Conclusion}

We show that recognizing community-valued contributions is a harder task than generic quality assessment---a finding that holds consistently across the four model families we evaluate on 15 communities. Value, or desirability, is fundamentally contextual: community-specific models consistently outperform global models, while the same linguistic features predict approval in opposite directions across community types. Fine-tuned SLMs and supervised encoders substantially outperform prompted LLMs, which exhibit a false-negative bias, discarding over 80\% of community-valued content under vanilla prompting. Conversational context narrows, but does not close, this gap, highlighting the importance of parametric adaptation to community-specific norms.
These findings point toward modular, community-adapted architectures as a path for feed curation, positive moderation, and reward modeling in community-aware systems.
\textsc{Vastu} provides a shared evaluation substrate for future work in this space.
\section*{Limitations}

\para{Approval labels and non-linguistic signals:} Our study uses Reddit \textit{score} as a proxy for community desirability. While this metric is a validated signal of collective approval, it remains an imperfect proxy subject to visibility bias, timing effects, and bandwagon dynamics that may conflate content quality with popularity. Our focus throughout is on evaluating model families based on language and content signals; we recognize that preference prediction also depends on non-linguistic factors such as comment timing, posting rank, and thread position. A more thorough investigation of how these structural signals interact with linguistic content is left for future work, as it requires purpose-built evaluation beyond the scope of this study.

\para{Platform and temporal scope:} Our analysis focuses exclusively on Reddit data from 2016--2017 due to open-sourced data availability. As a consequence, our findings reflect the specific demographics, platform norms, and community dynamics of that period and may not fully generalize to other social media environments or contemporary community behavior.

\para{Interpretability and causality:} While our SHAP and Chain-of-Thought analyses provide valuable interpretability, they identify correlations and self-reported reasoning rather than established causal drivers of community value. The features and reasoning patterns that predict approval may reflect surface regularities rather than the underlying norms communities actually hold.

\para{Task formulation:} We operationalize community value detection as binary classification, which captures preference at the extremes of the score distribution but does not directly evaluate listwise reordering in the information retrieval sense. Future work should connect these detectors to ranking and retrieval evaluations to assess their utility in downstream curation and recommendation settings. Similarly, since our comments are drawn from a relatively short, contiguous window of about 9 months, we did not structure the training and evaluation as a time-period based split where models learn to predict future values from from past data. However, future work could explore that setting with more longitudinal data available.

\para{Experimental and model coverage:} Due to budget constraints, we run a single evaluation seed for language models and reduce variability by setting temperature to 0, though some uncertainty may remain. We also leave community-specific LoRA fine-tuning and mixture-of-expert architectures for future exploration. More broadly, the rapid evolution of LLM capabilities means that newer architectures or agentic workflows could yield different performance trade-offs worth investigating.

\section*{Acknowledgments}

This research was supported by NSF CAREER IIS-2439433. 

This work used the Delta system at the National Center for Supercomputing Applications through allocation \#240481 from the Advanced Cyberinfrastructure Coordination Ecosystem: Services \& Support (ACCESS) program, which is supported by National Science Foundation grants \#2138259, \#2138286, \#2138307, \#2137603, and \#2138296.

\section*{AI Involvement Disclosure}

The research conducted in our work used artificial intelligence tools (e.g., Grammarly, ChatGPT) solely for improving the grammar and readability of certain sections of the manuscript, and for partial development of the codebase for the evaluations conducted through Claude Code, Codex, and Cursor. All scientific content, interpretations, and decisions reflect the original work, judgment and intellectual contributions of the research team. 
\section*{Ethical Considerations}

Our work uses publicly available Reddit data, which we handle in accordance with the platform's terms of service and privacy policies. All usernames and author identifiers are anonymized in our analysis. While our dataset focuses on highly-upvoted content, it may contain offensive language or controversial viewpoints reflective of diverse community norms. The authors do not endorse such content but analyze it to understand what communities reward in practice.

Our methods could potentially be misused by malicious actors to game recommendation systems or artificially inflate content visibility by reverse-engineering community preferences. Additionally, while value-aligned curation aims to improve user experience, overly optimized systems risk creating echo chambers or entrenching existing biases within communities. We emphasize that our models should be deployed with appropriate safeguards, transparency about ranking mechanisms, and ongoing monitoring for unintended consequences. Finally, platforms implementing such systems should provide users with meaningful control over curation algorithms, ensuring it serves user agency rather than replacing it.

\bibliography{references}

\appendix
\section{Appendix}
\setcounter{table}{0}
\setcounter{figure}{0}
\renewcommand{\thetable}{A\arabic{table}}
\renewcommand{\thefigure}{A\arabic{figure}}

\subsection{Feature Set Ablation}\label{app:feature-ablation}

\begin{figure}[ht]
    \centering
    \includegraphics[width=\linewidth]{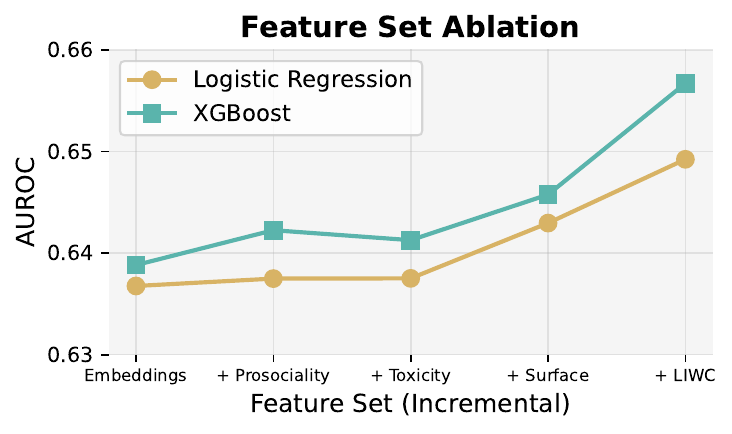}
    \caption{\textbf{Feature set ablation for feature-based models.} AUROC improves incrementally as feature sets are added, with LIWC providing the largest marginal gain.}
    \label{fig:rq1-feature-ablation}
\end{figure}

Before conducting computationally expensive encoder fine-tuning, we investigated which feature sets contribute most to detection performance. Using feature-based models (Logistic Regression and XGBoost), we evaluate five incremental configurations: (1) embeddings only, (2) + prosociality, (3) + toxicity, (4) + surface attributes, and (5) + LIWC. As shown in Figure~\ref{fig:rq1-feature-ablation}, performance improves monotonically as features are added, with LIWC categories providing the largest marginal gain (+0.01 AUROC for both models). All experiments in the main paper use the full feature set.

\subsection{Per-Topic Performance Breakdowns}\label{app:topic-breakdowns}

Tables~\ref{tab:rq1-topics} and~\ref{tab:llm-topics} report AUROC broken down by topic domain for feature-based/encoder models and language models respectively. 

From Table~\ref{tab:rq1-topics}, we observe meaningful variation in detection difficulty across topics. Politics communities show the highest predictability (0.76 AUROC for ModernBERT), possibly because valued content in these communities exhibits more consistent linguistic patterns (e.g., well-reasoned arguments, civility). Science and Gaming communities prove most challenging (0.64--0.72 AUROC), potentially due to the diverse nature of valued contributions across different scientific domains. Q\&A and Advice communities fall in between, with moderate predictability (0.68--0.75 AUROC). 

Similar patterns hold for the LLM and fine-tuned SLM predictions.

\begin{table*}[ht]
\small
\sffamily
\centering
\begin{minipage}[t]{0.48\textwidth}
    \vspace{0pt}
    \centering
    \resizebox{\linewidth}{!}{
    \begin{tabular}{lcccc}
    \toprule
    \textbf{Topic} & \textbf{LogReg} & \textbf{XGBoost} & \textbf{BERT} & \textbf{ModernBERT} \\
    \midrule
    Gaming & $0.67$ & $0.66$ & \cellcolor{badcell}$0.69$ & \cellcolor{badcell}$0.72$ \\
    Science & \cellcolor{badcell}$0.65$ & \cellcolor{badcell}$0.64$ & \cellcolor{badcell}$0.69$ & \cellcolor{badcell}$0.72$ \\
    Q\&A & $0.68$ & $0.68$ & $0.72$ & $0.75$ \\
    Advice & $0.68$ & $0.67$ & $0.71$ & $0.74$ \\
    Politics & \cellcolor{goodcell}$0.69$ & \cellcolor{goodcell}$0.69$ & \cellcolor{goodcell}$0.73$ & \cellcolor{goodcell}$0.76$ \\
    \bottomrule
    \end{tabular}}
    \captionof{table}{\textbf{Topic-level AUROC for feature-based and encoder models (local training).} Politics shows highest predictability; Science shows lowest across all architectures. Highlighted cells indicate per-column best and worst.}
    \label{tab:rq1-topics}
\end{minipage}
\hfill
\begin{minipage}[t]{0.48\textwidth}
    \vspace{0pt}
    \centering
    \resizebox{\linewidth}{!}{
    \begin{tabular}{lccccc}
    \toprule
    \textbf{Topic} & \textbf{4o-mini} & \textbf{4o} & \textbf{5-mini} & \textbf{Gemma-3} & \textbf{Qwen3} \\
    \midrule
    Gaming & \cellcolor{badcell}$0.53$ & $0.56$ & \cellcolor{badcell}$0.51$ & \cellcolor{badcell}$0.59$ & \cellcolor{badcell}$0.60$ \\
    Science & $0.55$ & \cellcolor{badcell}$0.55$ & \cellcolor{badcell}$0.51$ & $0.63$ & $0.62$ \\
    Q\&A & $0.57$ & $0.59$ & $0.54$ & $0.65$ & $0.65$ \\
    Advice & $0.56$ & $0.58$ & $0.54$ & $0.68$ & $0.65$ \\
    Politics & \cellcolor{goodcell}$0.59$ & \cellcolor{goodcell}$0.62$ & $0.54$ & \cellcolor{goodcell}$0.70$ & \cellcolor{goodcell}$0.68$ \\
    \bottomrule
    \end{tabular}}
    \captionof{table}{\textbf{Topic-level AUROC for language models.} Prompted LLMs use Zero-Shot; fine-tuned SLMs use LoRA. Politics communities show the highest predictability across all models, while Gaming proves most challenging, mirroring the pattern in Table~\ref{tab:rq1-topics}.}
    \label{tab:llm-topics}
\end{minipage}
\vspace{-8pt}
\end{table*}

\subsection{SHAP Analysis for Supervised Models}\label{app:shap-analysis}

\begin{figure*}[t]
    \centering

    \begin{minipage}[t]{0.48\textwidth}
        \vspace{0pt}
        \centering
        \includegraphics[width=\linewidth]{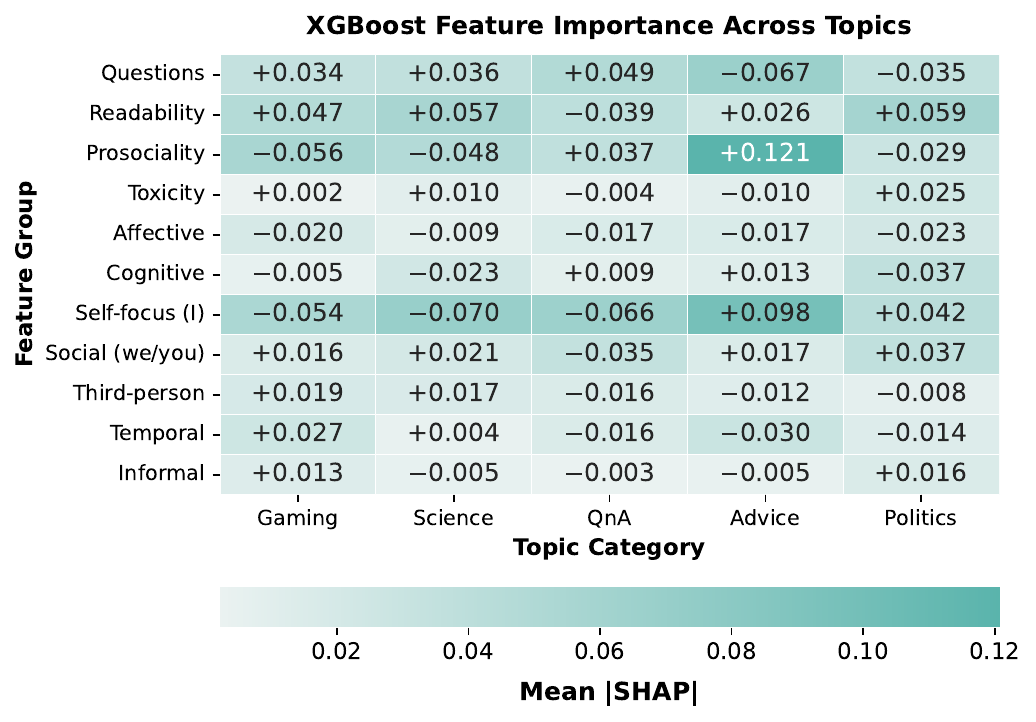}
        \caption{\textbf{SHAP feature importance for XGBoost across topics.} Cells show mean signed SHAP values aggregated across communities within each topic. Feature importance varies substantially by communities.}
        \label{fig:shap-heatmap}
    \end{minipage}
    \hfill
    \begin{minipage}[t]{0.48\textwidth}
        \vspace{0pt}
        \small
        \sffamily
        \centering
        \resizebox{\linewidth}{!}{
        \begin{tabular}{ll}
        \toprule
        \textbf{Feature Group} & \textbf{Constituent Features} \\
        \midrule
        \rowcollight Prosociality & agreement, politeness, support \\
        Toxicity & toxicity \\
        \rowcollight Affective & vader compound, posemo, negemo \\
        Readability & flesch reading ease \\
        \rowcollight Questions & interrogative fraction, interrog \\
        Self-focus (I) & i \\
        \rowcollight Social (we/you) & we, you, social \\
        Cognitive & cogproc, insight, cause, certain, tentat \\
        \rowcollight Temporal & focuspast, focuspresent, focusfuture, time \\
        Informal & informal, swear, netspeak \\
        \rowcollight Third-person & shehe, they \\
        \bottomrule
        \end{tabular}}
        \captionof{table}{Mapping of individual features to interpretable feature groups used in SHAP analysis.\vspace{-16pt}}
        \label{tab:feature-groups}
    \end{minipage}
\end{figure*}

To understand what signals supervised models rely on, we compute SHAP values~\cite{lundberg2017unified} for feature importance analysis. We focus on XGBoost as it achieves the strongest performance among feature-based models and produces stable importance estimates. While SHAP can be applied to transformer models, their learned internal representations make interpretation less straightforward, and we therefore restrict this analysis to feature-based approaches where input features map directly to interpretable constructs. To aid interpretation, we aggregate our 80+ individual features into 11 semantically coherent groups (e.g., \textit{Prosociality} combines agreement, politeness, and support scores; \textit{Cognitive} combines LIWC categories for cognitive processes, insight, causation, certainty, and tentativeness). Full groupings are provided in \autoref{tab:feature-groups}. For each topic category, we compute mean signed SHAP values across communities and test instances, where positive values indicate the feature group pushes predictions toward ``valued'' and negative values push toward ``not valued.''

\para{Feature importance varies substantially across community types:}
From Figure~\ref{fig:shap-heatmap}, we observe the heterogeneity in what signals predict valued content across topic categories. The most notable example is \textit{Prosociality}: in Advice communities, prosocial language is the single strongest predictor of valued content (+0.121 mean SHAP), yet the same features show negative or negligible effects in Gaming (-0.056), Science (-0.048), and Politics (-0.029). Similarly, \textit{Self-focus} (first-person singular pronouns) strongly predicts valued content in Advice (+0.098)---where personal experience is central---but is penalized in Science (-0.070) and Q\&A (-0.066) communities that prioritize objective, generalizable information. This pattern reinforces our RQ1 finding that community-specific models outperform global ones: a model learning that prosociality universally predicts value would systematically misclassify content in Gaming and Politics communities.

\para{Question-asking and readability show interpretable community-specific patterns:}
Beyond prosociality, other feature groups exhibit patterns that align with community purposes. \textit{Questions} (interrogative framing) positively predict valued content in Q\&A (+0.049) and Science (+0.036)---communities oriented around inquiry---but negatively predict value in Advice (-0.067) and Politics (-0.035), where users seek definitive guidance or argumentation rather than open questions. \textit{Readability} shows broadly positive effects, particularly in Politics (+0.059) and Science (+0.057), suggesting that clear writing is valued in communities dealing with complex topics. Interestingly, \textit{Cognitive} language (reasoning, causation, certainty) is penalized in Politics (-0.037) and Science (-0.023) but weakly positive in Advice (+0.013), perhaps because overly analytical framing reads as hedging in debate-oriented communities but signals thoughtfulness in support contexts. These interpretable patterns suggest that supervised models learn genuine community norms rather than spurious correlations in data.

\para{Robustness of SHAP-derived patterns:} To verify that the SHAP feature-group patterns are stable rather than seed- or split-dependent, we quantify stability along three axes: \textbf{(1) Cross-fold rank stability:} across the 5 CV folds, the mean pairwise Spearman correlation of feature-group importance rankings is $\bar{\rho} = 0.86$ (min $0.79$); \textbf{(2) Cross-subreddit rank stability:} within each topic, the mean pairwise Spearman across the three constituent subreddits is $\bar{\rho} = 0.71$, indicating that group-level orderings are largely shared within a topic while retaining community-specific variation; and \textbf{(3) Top-k stability:} the top-5 feature groups have a mean cross-fold Jaccard overlap of 0.82, with a coefficient of variation of of 0.11. Overall, these show that the SHAP analysis is fairly stable with no changes to our interpretation, and readers of our work can get better insights into how much they can rely on the analysis accordingly.

\subsection{Extended Results on LLM Error Analysis}\label{app:llm-error-coding}

\autoref{tab:rq3-error-taxonomy} presents the entire set of six failure modes identified and presented in \S\ref{sec:error-analysis}. We additionally present a comparative analysis of the interpretive patterns we observe from the thematic analysis in comparison to the SHAP analysis in \S\ref{app:shap-analysis}.

\para{Feature-based and LLM signals are complementary but not identical:}
Comparing SHAP patterns to CoT themes highlights a key difference in \emph{representation}. Feature-based models rely on measurable proxies (e.g., prosociality, interrogative style, self-focus, readability), which produce clear community-conditional effects (Figure~\ref{fig:shap-heatmap}). In contrast, LLM explanations operate at a more semantic and normative level---explicitly reasoning about \emph{why} a community might value a contribution (e.g., ``evidence-based,'' ``on-topic,'' ``actionable,'' ``constructive'')---but this abstraction flattens community nuance. In several incorrect cases (analyzed in RQ3), LLMs over-apply these generic heuristics even when those styles are rewarded in that community based on the SHAP analysis. Together, these results suggest that supervised feature-based models capture stable, interpretable correlates of community value, while LLM explanations reflect a higher-level normative theory that is often miscalibrated.

\begin{table*}[t]
\small
\sffamily
\centering
\resizebox{0.85\linewidth}{!}{
\begin{tabular}{lp{8.5cm}c}
\textbf{Type} & \textbf{What Typically Goes Wrong} & \textbf{Type} \\
\midrule
\rowcollight \textbf{T1: Wrong community norm} & Misunderstands what subreddits typically rewards (including community jargon), so comment is judged as ``not a good fit.'' & FN \\
\textbf{T2: ``More effort is better''} & Treats short / punchy / obvious comments as low-quality, even when that style is commonly upvoted. & FN \\
\rowcollight \textbf{T3: Missing thread context} & Judges replies as ``unclear'' because they rely on the quotes or shared context that is not present in the standalone text. & FN \\
\textbf{T4: Evidence bar mismatch} & Applies the wrong standard for evidence (too strict or too loose), especially in Science and Q\&A communities. & FN / FP \\
\rowcollight \textbf{T5: Humor / sarcasm} & Misreads jokes, irony, or sarcasm (takes it literally or dismisses it), leading to mismatched value judgments. & FN / FP \\
\textbf{T6: Profanity penalty} & Over-penalizes profanity even when it is not targeted harassment and is common/acceptable in the community. & FN \\
\bottomrule
\end{tabular}}
\caption{\textbf{LLM Error Taxonomy.} We summarize failure patterns in LLM misclassifications.\vspace{-12pt}}
\label{tab:rq3-error-taxonomy}
\end{table*}

\subsection{Threshold Sensitivity Analysis}\label{app:threshold-sensitivity}

Our primary experiments define \texttt{HIGH} as the top 5\% of subreddit-specific scores ($q{=}0.95$) and \texttt{LOW} as the bottom 30\% ($q{=}0.70$). To assess whether model rankings and key conclusions depend on this choice, we re-evaluate one representative model from each paradigm at two alternative \texttt{HIGH} thresholds---a softer cutoff ($q{=}0.90$) and a stricter cutoff ($q{=}0.99$)---holding all other conditions fixed.

\begin{table}[ht]
\small
\sffamily
\centering
\begin{tabular}{lccc}
\textbf{Model + Paradigm} & $q{=}0.90$ & $q{=}0.95$ & $q{=}0.99$ \\
\midrule
\rowcollight GPT-4o Zero-Shot & $0.57$ & $0.60$ & $0.64$ \\
Gemma-3-4B LoRA & $0.62$ & $0.65$ & $0.70$ \\
\rowcollight LogReg Local & $0.66$ & $0.69$ & $0.74$ \\
ModernBERT Local & $0.70$ & $0.74$ & $0.79$ \\
\midrule
Spearman $\rho$ vs.\ $q{=}0.95$ & $1.00$ & --- & $1.00$ \\
\bottomrule
\end{tabular}
\caption{\textbf{Threshold sensitivity across paradigms.} AUROC at three \texttt{HIGH} label thresholds for one representative model per paradigm. Spearman $\rho$ measures rank correlation of model AUROC scores relative to the defaultd.}
\label{tab:threshold-sensitivity}
\end{table}

Table~\ref{tab:threshold-sensitivity} shows that absolute AUROC rises with threshold stringency: at $q{=}0.99$, the most highly upvoted comments are more linguistically distinctive, yielding a more separable classification problem. The shift is roughly uniform across paradigms (${\approx}{+}0.05$ from $q{=}0.95$ to $q{=}0.99$, ${\approx}{-}0.03$--$0.04$ from $q{=}0.95$ to $q{=}0.90$), so no paradigm benefits disproportionately from a stricter label. Crucially, model rankings are fully preserved: ModernBERT $>$ LogReg $>$ Gemma-3-4B $>$ GPT-4o at every threshold, with Spearman $\rho = 1.00$ between each alternative and the default. These results confirm that the adaptation-versus-prompting gap reported in the main text is not an artifact of the specific approval cutoff chosen to define community-valued content.

\subsection{Transformer-Based Model Architecture}\label{app:transformer-architecture}

\begin{figure}[ht]
    \centering
    \includegraphics[width=\linewidth]{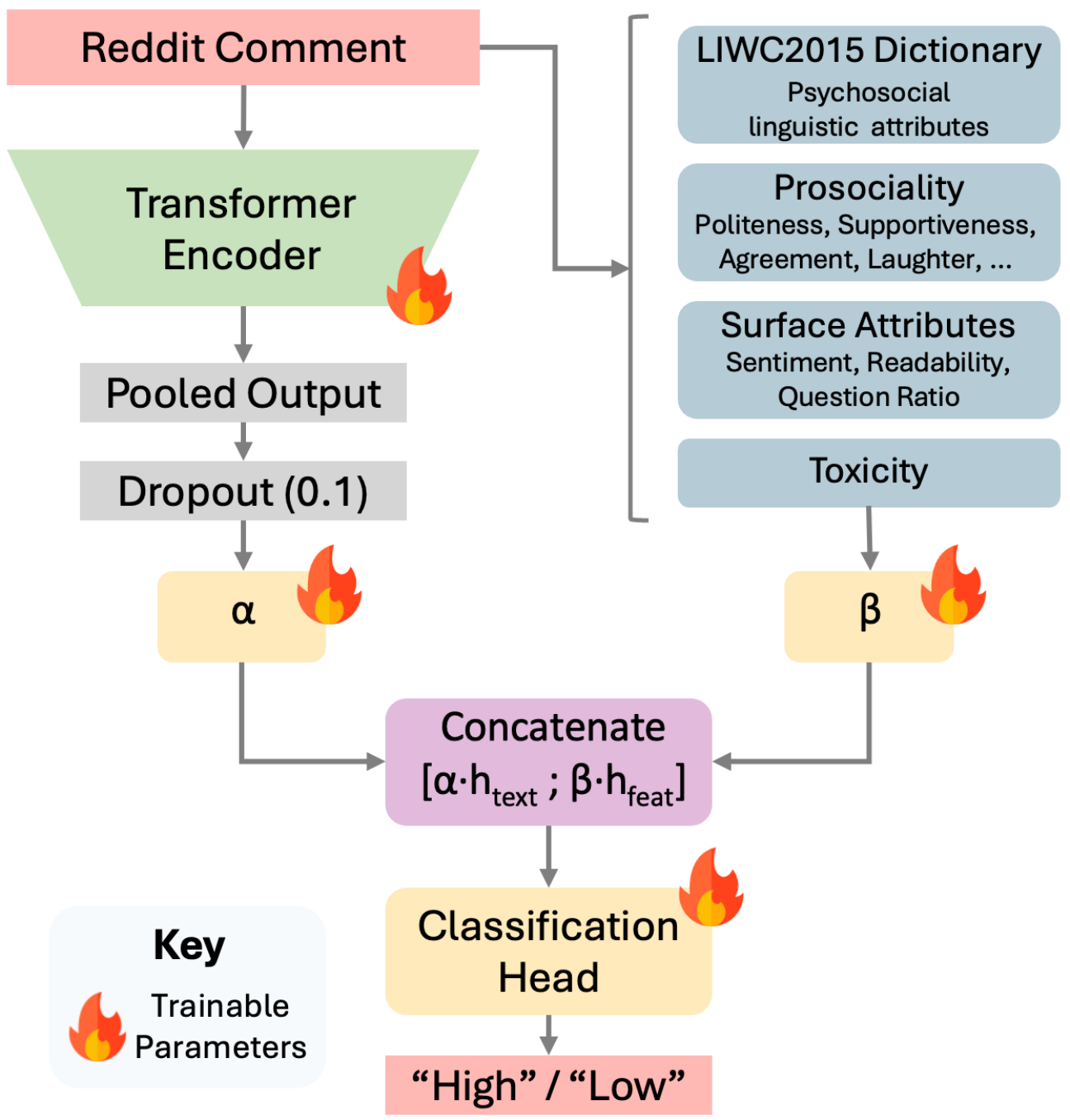}
    \caption{\textbf{Architecture of our transformer-based feature fusion model.} The input Reddit comment is processed through a transformer encoder to obtain contextual representations, while linguistic features are extracted in parallel. Learnable weights allow the model to learn the optimal contribution of text embeddings versus linguistic features. The concatenated representation is passed through a classification head to predict whether the comment will be highly-valued by the community.}
    \label{fig:transormer-architecture}
\end{figure}

Figure~\ref{fig:transormer-architecture} shows a visual depiction of the transformer architecure we use for modeling community-valued content in our work with supervised encoders.

\subsection{Prompting Strategies for Large Language Models}\label{app:llm-prompts}

We evaluate four distinct prompting strategies for LLM-based detection of community-valued content. Below we provide the complete prompts used for each condition. All prompts are shown for the \texttt{comment+community} context, with addition of the preceding comment for the \texttt{comment+community+parent} condition, and removal of the \texttt{community} context for the \texttt{comment-only} condition.

\subsubsection{Condition 1: Zero-Shot Vanilla}
The baseline condition with minimal instruction.
\paragraph{System Prompt:}
\begin{quote}\small
\textit{You are an expert at understanding online community dynamics and what content resonates with different communities.}
\end{quote}

\paragraph{User Prompt Template:}
\begin{quote}\small
\textit{You are evaluating whether a comment would be highly upvoted in the Reddit community r/{subreddit\_name}.}

\textit{\textbf{Community Description:} {subreddit\_description}}

\textit{\textbf{Comment to Evaluate:} {target\_comment}}

\textit{Would this comment likely receive high approval (top $5\%$ of upvotes) from this community?}

\textit{Respond with exactly one of: HIGH or LOW}
\end{quote}

\subsubsection{Condition 2: Chain-of-Thought (CoT)}
Adds explicit reasoning instruction to elicit step-by-step analysis.
\paragraph{System Prompt:}
\begin{quote}\small
\textit{You are an expert at understanding online community dynamics and what content resonates with different communities. You think carefully and systematically about what makes content valuable to specific communities.}
\end{quote}
\paragraph{User Prompt Template:}
\begin{quote}\small
\textit{You are evaluating whether a comment would be highly upvoted in the Reddit community r/{subreddit\_name}.}

\textit{\textbf{Community Description:} {subreddit\_description}}

\textit{\textbf{Comment to Evaluate:} {target\_comment}}

\textit{Think step-by-step:}

\begin{enumerate}
\item \textit{What is this community about and what do they typically value?}
\item \textit{What qualities does the target comment have that this community might appreciate or dislike?}
\item \textit{Based on this analysis, would this comment likely receive high approval (top $5\%$ of upvotes) from this community?}
\end{enumerate}

\textit{After your reasoning, provide your final prediction.}

\textit{Format your response as:}

\textit{PREDICTION: [HIGH or LOW]}

\textit{REASONING: [BRIEF step-by-step analysis]}
\end{quote}

\subsubsection{Condition 3: Few-Shot (k=3)}
Provides three labeled examples from the same community before evaluation. 

\paragraph{System Prompt:}
\begin{quote}\small
\textit{You are an expert at understanding online community dynamics and what content resonates with different communities.}
\end{quote}

\paragraph{User Prompt Template:}
\begin{quote}\small
\textit{You are evaluating whether a comment would be highly upvoted in the Reddit community r/{subreddit\_name}.}

\textit{\textbf{Community Description:} {subreddit\_description}}

\textit{Here are some examples of comments from this community and whether they received high approval:}

\textit{\textbf{Example 1:}}

\textit{Comment: {example1\_comment}}

\textit{Approval: {example1\_label}}

\textit{\textbf{Example 2:}}

\textit{Comment: {example2\_comment}}

\textit{Approval: {example2\_label}}

\textit{\textbf{Example 3:}}

\textit{Comment: {example3\_comment}}

\textit{Approval: {example3\_label}}

\textit{Now evaluate this comment:}

\textit{\textbf{Comment to Evaluate:} 
{target\_comment}}

\textit{Would this comment likely receive high approval (top $5\%$ of upvotes) from this community?}

\textit{Respond with exactly one of: HIGH or LOW}
\end{quote}

\subsubsection{Condition 4: Value-Augmented Prompting}

Provides the Schwartz Basic Human Values framework~\cite{Schwartz1992UniversalsIT,Schwartz2012RefiningTT} as an analytical lens for evaluation, based on the framework utilized by \citet{2026whosevalues}.

\paragraph{System Prompt:}
\begin{quote}\small
\textit{You are an expert at understanding online community dynamics and human values. You use established psychological frameworks to analyze what content resonates with different communities.}
\end{quote}

\paragraph{User Prompt Template:}
\begin{quote}\small
\textit{You are evaluating whether a comment would be highly upvoted in the Reddit community r/{subreddit\_name}.}

\textit{\textbf{Community Description:} {subreddit\_description}}

\textit{To help with your evaluation, consider the Schwartz Basic Human Values framework. These are universal values that motivate human behavior:}
\begin{enumerate}
\item \textit{\textbf{Self-Direction}: Independent thought and action (creativity, freedom, curiosity)}
\item \textit{\textbf{Stimulation}: Excitement, novelty, and challenge}
\item \textit{\textbf{Hedonism}: Pleasure and sensuous gratification}
\item \textit{\textbf{Achievement}: Personal success through demonstrating competence}
\item \textit{\textbf{Power}: Social status, prestige, control over resources}
\item \textit{\textbf{Security}: Safety, harmony, stability of society and relationships}
\item \textit{\textbf{Conformity}: Restraint of actions that might harm others or violate norms}
\item \textit{\textbf{Tradition}: Respect and commitment to cultural/religious customs}
\item \textit{\textbf{Benevolence}: Concern for the welfare of close others}
\item \textit{\textbf{Universalism}: Understanding, tolerance, and protection of all people and nature}
\end{enumerate}

\textit{\textbf{Comment to Evaluate:} {target\_comment}}

\textit{Consider:}
\begin{itemize}
\item \textit{Which values does this comment express or appeal to?}
\item \textit{Which values are likely important to this specific community?}
\item \textit{Does the comment align with or violate values this community cares about?}
\end{itemize}

\textit{Based on this value-based analysis, would this comment likely receive high approval (top 5\% of upvotes) from this community?}

\textit{Format your response as:}

\textit{PREDICTION: [HIGH or LOW]}

\textit{VALUES\_EXPRESSED: [List the 1-3 most relevant Schwartz values this comment expresses]}

\textit{COMMUNITY\_VALUES: [List the 1-3 Schwartz values you think this community prioritizes]}

\textit{ALIGNMENT: [Brief assessment of alignment]}
\end{quote}

\section{Computational Resources}\label{app:compute-resources}

All experiments using open-source models were run on organization GPU servers equipped with 3xNVIDIA A40. The API costs for running OpenAI models were around $\$250$.

\end{document}